\documentstyle[12pt]{article}
	 \makeatletter
	 \def\imi{|I_iM_i>}
	 
	 \def\ifmf{<I_fM_f|}
	 
	 \def\e0{\hat{\bf e}_0}
	 
	 \def\bgq{\begin{equation}}
	 \def\edq{\end{equation}}
	 \def\bga{\begin{eqnarray}}
	 \def\eda{\end{eqnarray}}
	 \def\be11{^{11}Be}
	 \def\halfp{{1\over 2}^+}
	 \def\halfm{{1\over 2}^-}
	 \def\pb{^{208}Pb}
	 \def\lar{\longrightarrow}
	 \def\thefigure{\@arabic\c@figure}\def\fps@figure{tbp}
	 \def\ftype@figure{1}\def\ext@figure{lof}
	 \def\fnum@figure{\protect\footnotesize Fig.\ \thefigure}
	 \def\thetable{\@arabic\c@table}
	 \def\fps@table{tbp}\def\ftype@table{2}\def\ext@table{lot}
	 \def\fnum@table{\protect\footnotesize Table \thetable}
	 \makeatother
\begin{document}
\vspace*{0.3in}
\begin{center}
  {\Large \bf A Coupled-Channels Study of \\
   $^{11}Be$ Coulomb Excitation }\\
  \bigskip
  \bigskip
  {\Large
  C.A. Bertulani$^\dagger$, L.F. Canto} \\
  \bigskip
  Instituto de F\'\i sica, Universidade Federal do Rio de Janeiro\\
  Cx. Postal 68528\\
  21945-970 Rio de Janeiro, RJ, Brazil\\
  \bigskip
  and\\
  \bigskip
  {\Large M.S. Hussein$^+$}  \\
\bigskip

		    Center for Theoretical Physics\\
		    Laboratory for Nuclear Science and Department of Physics \\
		 Massachusetts Institute of Technology, \
	       Cambridge, Massachusetts, 02139,  USA $^*$\\
				and\\
	  Institute for Theoretical Atomic and Molecular Physics\\
	   at the Harvard-Smithsonian Center for Astrophysics \\
	   60 Graden St. Cambridge, Massachusetts, 02138, USA $^\ddagger$\\

 \bigskip
\end{center}

\bigskip
\centerline{\bf ABSTRACT}
\begin{quotation}
\vspace{-0.10in}
We study the effects of channel coupling in the excitation of
$^{11}Be$ projectiles incident on heavy targets.  The
contribution to the excitation from the Coulomb and the nuclear
fields in peripheral collisions are considered.  Our results are
compared with recent data on the excitation of the
$\halfm$ state in $^{11}Be$ projectiles.
We show that the experimental
results cannot be explained, unless very unusual parameters are
used.

\end{quotation}
\vfil
\noindent $\dagger$ E-mail: bertu@if.ufrj.br
\medskip

\noindent + Permanent address: Instituto de F\'\i sica,
Universidade de S\~ao Paulo, Cx. Postal 20516, 01498 S\~ao
Paulo, SP, Brazil. Supported in part by FAPESP.
\medskip

\noindent * Supported
in part by funds provided by the U.S. Department of Energy
(D.O.E.) under cooperative agreement
DE-FC02-94ER40818
\medskip

\noindent $\ddagger$ Supported by the National Science
Foundation

\newpage
\baselineskip 4ex

Unstable nuclei  are often studied in reactions induced by
secondary beams. Examples of such reactions include
elastic scattering, fragmentation and Coulomb
excitation in collisions with very heavy targets \cite{PG}.
Coulomb excitation is especially useful, since the interaction mechanism
is well known. The cross sections for Coulomb excitation of radioactive
beams yield precious information on the intrinsic electromagnetic
moments of these nuclei. Such information are hard to obtain with
other methods due to the short lifetime of unstable projectiles.
Recently, the Coulomb excitation of the halo nucleus $^{11}Be$ has been
investigated \cite{nakamura,anne}. In ref. \cite{nakamura}
the transition from the ground state (a parity-inverted $\halfp$ state)
of $\be11$ to the continuum (with threshold at 504 keV)
was studied and a good agreement with the theory was found. On the
other hand, in ref. \cite{anne} the Coulomb excitation of the
$\halfm$ state at 320 keV (the  only excited state in
$\be11$ \cite{millener}) was studied in collisions of 45 MeV.A $\be11$ beams on
$\pb$ targets.
Amazingly, it was found that
the measured cross section, 191$\pm$26 mb, is a factor
2 smaller than that expected from first order
perturbation theory,
using B(E1)=0.116$\pm$0.012 e$^2$ fm$^2$.
This B(E1)-value is an average over three distinct experiments
which yield the
lifetime of 166$\pm$15 fs for this state.

It has been suggested that the reason for the above mentioned discrepancy
could be the coupling of the
bound states in $\be11$ and the continuum, or the
contribution from nuclear excitation causing a destructive Coulomb-nuclear
interference
\cite{anne,hansen}.
We will investigate these possibilities here using a
semiclassical coupled-channels approach to
the Coulomb and nuclear excitation of the $\halfm$ state in $\be11$.
The spin and parities
of the states involved imply that the Coulomb
dipole excitation corresponds to
the largest contribution to the cross section. This has been
experimentally observed \cite{nakamura,anne} and is theoretically
understood \cite{esbensen}.

We treat the Coulomb excitation in the semiclassical formalism,
assuming a straight-line trajectory for the projectile, and including
relativistic effects in the interaction. The matrix element for the
Coulomb dipole potential in a collision with impact parameter $b$ is
\bga
<k|V_{(E1)}(t)|i>&=& \sqrt{2\pi\over 3}\ \gamma \
\Bigg\{ {\cal E}_1 (\tau) \ \Big[ {\cal M}_{ki} (E1, -1) -
{\cal M}_{ki} (E1, 1)\Big]
\nonumber \\
&+&\sqrt{2} \ \gamma \ \tau\ \bigg[ {\cal E}_1(\tau)
-
i{\omega v b\over c^2}
\ \Big( 1+\tau^2 \Big)\  {\cal E}_2(\tau)
\bigg] \ {\cal M}_{ki} (E1, 0) \Bigg\}
\ ,
\eda
where $\gamma$ is the Lorentz factor, $\tau=(\gamma v/b)\ t$, $\hbar \omega$
is the excitation energy,
and
\bgq
{\cal E}_1(\tau)={ Z_T e  \over
b^2 \ [1+\tau^2]^{3/2}} \
\ \ \ \
{\rm and} \  \ \
{\cal E}_2 (\tau)={ Z_T e \tau  \over b \ [1+\tau^2]^{3/2}}
\edq
are respectively  the transverse and longitudinal electric fields
generated by the target nucleus.
The matrix elements for electric dipole excitations are given by
${\cal M}_{ki}(E1,m)=e\ <k|rY_{1m}|i>$, where $e$ is the electron charge.

We will solve numerically the coupled-channels equations
\bgq
{d a_k\over dt} (t) = \sum_i <k|V(t)|i> \
\exp\Big\{i (E_k-E_i) t/\hbar\Big\} \
a_i (t)
\ ,
\edq
where $|k>$ denote the (discrete) nuclear states.
For the Coulomb
coupling, the Wigner-Eckart theorem allows  us to write the matrix
elements $<k|V(t)|i> $ in terms of the
$B(E;M\lambda)$-values for the electromagnetic multipole
transitions. For the transition $\halfp \lar \halfm$ we use the
previously mentioned B(E1)-value.
The Coulomb dipole transitions to the continuum
are treated by means of a discretization procedure. In refs.
\cite{nakamura,esbensen} it was shown that the dipole response
for the transitions from the bound-state to the continuum
can be appreciably well
accounted for by neglecting final state interactions and using
the asymptotic value of the ground state wave function, represented by
an Yukawa tail. This dipole response is given by \cite{PG}
\bgq
{1\over e^2} \ { dB(E1; \ E_x)\over dE_x} = C \ {\sqrt{S} \ (E_x-S)^{3/2}
\over E_x^4} \ ,
\edq
where $E_x$ is the excitation energy, and S is the separation energy
of the valence neutron in $\be11$. $C$ is a normalizing constant, which
the two-body
($n+^{10}Be$) model predicts to be independent of the separation
energy.
In the experiment of ref. \cite{nakamura} it was found that the best
fit to the data corresponds to $C=3.73\pm 0.7$ fm$^2$/MeV. We also use
this value for the transition $\halfm \lar continuum$.

The continuum is discretized so that the B(E1)-values from the
bound states to the
$n^{th}$ state in the continuum are given by
$$B(E1; \ i \lar n)=
\Delta E_x \cdot {dB(E1,\ E_x)\over dE_x}\Bigg|_{E_x=E_n}\ ,$$
which can be calculated with help of eq.   (4).
Above, $\Delta E_x$ is the spacing in the continuum energy mesh.
In our numerics
we use $\Delta E_x=0.3$ MeV, and a total of 10 discretized continuum
states. This mesh covers the most important part of the continuum
dipole response function in $\be11$ \cite{nakamura}.
A  phase convention for the nuclear states
can be found so that the reduced matrix elements
$\ifmf | {\cal M} (E;M\lambda) |\imi$ are
real \cite{AW65}. We noticed that in the present problem
the sign of the matrix elements do not  appreciably affect
the results. We then set all matrix elements as positive.

With the above described  procedure,
the matrix elements for the transitions
from the bound states to the continuum are fixed. We neglect
the coupling between the continuum states, since
it corresponds to reacceleration effects, which has been
shown to be small for this  system \cite{esbensen}.
The integrated dipole response in the continuum is obtained from
eq. (4) as
\bgq
B(E1; i \lar cont.)/e^2= {\pi C\over 16 S} \ ,
\edq
where $i$ stands for one of the two bound states in $\be11$.
Using the experimental value of $C$, we get
$B(E1; \halfp \lar cont.)=1.45$  $e^2$ fm$^2$,  for the ground state,
and $B(E1; \halfm \lar cont.)=4.06$  $e^2$ fm$^2$
for the first excited state.

With respect to the nuclear interaction  in peripheral
collisions,
we expect that the most relevant contributions arise from the
monopole and quadrupole isoscalar excitation modes. Isovector excitations
are strongly suppressed
\cite{Sa87} due to the approximate
charge independence of the nuclear interaction. We adopt the
folding optical potential of ref. \cite{sagawa}.
For the ground state density of $^{11}Be$ we use the results \cite{sagawa}
of the Hartree-Fock formalism with Skyrme interaction, and for $\pb$
we take a Fermi density
with radius $R=6.67$ fm and diffuseness $a=0.55$ fm.
The monopole and quadrupole transition potentials were calculated with the
Tassie model, as explained in ref. \cite{sagawa}. In terms of the
optical potential $U_{opt}(r)$, they can be written
\bgq
V_N (r)=
\left\{ \begin{array}{ll}
\alpha_0 \Big[ 3 U_{opt} (r) + r \ d U_{opt} (r) / dr \Big] \ ,
    & \mbox{for monopole; }\\
    & \\
(\delta_2/\sqrt{5}) \ \Big[ dU_{opt} (r) / dr\Big] \ ,
    & \mbox{for quadrupole }
\end{array}
\right.
\edq
where $\alpha_0$ and $\delta_2$ are
parameters  to fit inelastic scattering data.
Since there are no such data on
$\be11$, we arbitrarily choose
$\alpha_0=0.1$ and $\delta_2=1$ fm. These values
correspond to  about 5\% of the energy-weighted sum rule, if
a state at 1 MeV excitation energy is assumed, and should be
reasonable for a qualitative calculation.
The nuclear couplings are given a time-dependence
though the application of
a Lorentz boost on the system. This amounts to
multiplying eq. (6) by the Lorentz factor $\gamma$, and using
$r=\sqrt{b^2+\gamma^2v^2t^2}$.

The cross section to excite the state $k$ are calculated
from the relation
\bgq
\sigma_k=2\pi\ \int db \ b\ \Big| a_k(b)
\Big|^2 \ \exp \Big( (2/\hbar v)\ {\rm Im} \big\{ \int dz \ U_{opt}(r)\big\}
\Big)
\ ,
\edq
where ${\bf r}=({\bf b}, \ z)$.
The exponential term accounts for the
strong absorption along the classical trajectory.

In table 1 we present the results of our calculations.
In the column ``Theory (2)",
reorientation effects caused by the magnetic dipole transitions
$\halfp\lar \halfp$ are included. For this purpose, we use
the Schmidt value
$B(M1, \halfp \lar \halfp ) = 0.087$ $e^2$ fm$^2$, and calculate
the magnetic dipole coupling through the same
procedure as that employed for $V_{E1}$ \cite{bertu}.
We observe that this effect can be neglected, since it
causes a negligible change in the cross sections.
On the other hand, the inclusion of the coupling to the
continuum yields more sizeable effects.
The cross section for the excitation of
the $\halfm$ state decreases by about 4\%.
This reduction is, however, still too
small to explain the discrepancy between experiment and
theory.
Finally, in the column ``Theory (4)",
we present effects of nuclear excitation. These
effects are also very small. The reason is that
the nuclear interaction
is limited to a very small impact parameter region, around
the grazing value, as illustrated in figure 1. The cross
sections for the nuclear excitation of monopole and
quadrupole states are respectively 7.07 mb and 6.22 mb.
A second reason for this fact is that Coulomb-nuclear
interference
only appears for high-order transitions, i.e., those involving
many excitation steps. This occurs because the Coulomb coupling
is dominated by the dipole term while the nuclear
coupling is dominated
by monopole  and quadrupole.
In table 1 we also show the dissociation cross section,
$\sigma_{cont.}$. It is of the same magnitude as
$\sigma_{1/2^-}$.

The above discussion indicates
that the most important factor leading to the
reduction of $\sigma_{1/2^-}$ is the coupling
between the bound states and the continuum.
We cannot be sure that the adopted
$B(E1; \halfm \lar cont.)$-value
is accurate, since the form of eq. (6) has been
shown to be appropriated
for transitions from the ground state
\cite{nakamura,esbensen}.
It is therefore worthwhile to let the strength
$B(E1; \halfm \lar cont.)$ vary and study the effects
on the
$\sigma_{1/2^-}$ cross section. This is shown
in figure 2, where we plot the quantity
\bgq
\Delta = {\sigma_{\halfm}(\xi) -\sigma_{\halfm}(\xi=1) \over
\sigma_{\halfm}(\xi=1)}\ ,
\edq
where $\xi$ is a
renormalization factor multiplying the strength given by
eq. (6).
We see that, increasing this strength  reduces the cross section
$\sigma_{1/2^-}$. However, this reduction
is much smaller than the factor 2 needed for an agreement
with the data.
We point out that, even assuming the complete exhaustion of the
energy-weighted dipole sum-rule for this state
(corresponding to the
indicated $\xi$-value in fig. 2), the reduction
still falls very short.

We conclude that taking into account
nuclear coupling and  multistep coupled-channel processes
cannot improve appreciably
the disagreement
between the experiment of ref. \cite{anne} and the theory.
The depopulation of the state
$\halfm$ arising from its coupling to the continuum, is not
large enough.
This result is not surprising, since the $\halfp\lar \halfm$
transition probability is very small, even for
grazing collisions (see fig. 1). Thus, first order
perturbation theory is
appropriate to calculate the cross sections.
A further increase of the
transition probability from the excited state to the continuum
would occur in the presence of a resonance in $\be11$
close to the
threshold. However, presently, this hypothesis lacks
experimental evidence
\cite{nakamura}. Further studies are needed
to clarify this matter.

\bigskip\bigskip

{\bf Acknowledgements}

We have benefited from useful discussions with H. Emling
and P.G. Hansen.
\bigskip\bigskip

\newpage
\noindent
{\bf Figure Captions}  \\
\begin{description}
\item[Fig. 1]
Excitation probabilities of the $\halfm$ state by the Coulomb field,
and of hypothetical low-lying monopole, and quadrupole, states by means
of the nuclear interaction, as a function of the impact parameter $b$.
The reaction $\be11+\pb$ at 45 MeV.A is considered.

\item[Fig. 2]
The ratio $
\Delta = \Big[\sigma_{\halfm}(\xi) -\sigma_{\halfm}(\xi=1)\Big] /
\sigma_{\halfm}(\xi=1)$ for the excitation of the $\halfm$
state in $\be11$ is shown as a function of the strength parameter
$\xi$.
\end{description}

\vskip 2cm
\noindent {\bf Table Caption}\\
\begin{description}
\item[Table 1 :]
Experimental and theoretical cross sections for
$^{11}Be $ excitation in the
$\be11\ (45 MeV.A)+^{208}Pb$ reaction.
In ``Theory (1)" the coupled-channel calculation was restricted
to the $\halfp$ and $\halfm$ channels.
In ``Theory (2)" the reorientation effect caused by the
M1-transition $\halfp \lar \halfp$ was introduced.
In ``Theory (3)"
we included the coupling between the
$\halfp$ and $\halfm$ states with the
continuum. Finally, in ``Theory (4)" we included
nuclear coupling effects.
\end{description}
\bigskip\bigskip
\newpage
\centerline {Table I}
\bigskip

\begin{center}
\begin{tabular}{|l|l|l||l|l|l|l|l|l|r|} \hline\hline
&Exp.&Theory (1)& Theory (2)&Theory (3)& Theory (4)\\
\hline\hline
$\sigma_{1/2^-}$\ (mb)&$191\pm 26$&405.81&405.80&388.2&390.2\\
\hline
$\sigma_{cont.}$\ (mb)&--&--&--&334.5&335.2\\
\hline\hline
\end{tabular}
\end{center}

\end{document}

\end{document}

